\documentclass[journal=jpclcd,manuscript=letter]{achemso}
\usepackage[version=3]{mhchem} 
\usepackage{amsmath,amsfonts,amssymb,siunitx}
\DeclareUnicodeCharacter{2212}{-}
\DeclareUnicodeCharacter{2009}{}
\usepackage{graphicx}


\author{Anita Girelli}
\email{anita.girelli@fysik.su.se}

\author{Iason Andronis}
\author{Aigerim Karina} 
\author{Sampad Bag} 
\affiliation[ Stockholm University]{Department of Physics, AlbaNova University Center, Stockholm University, 10691 Stockholm, Sweden}
\author{Lennart Bergström} 
\affiliation[ Stockholm University-chem]{  Department of Chemistry, Stockholm University, 10691 Stockholm, Sweden}
\author{Tomás S. Plivelic} 
\affiliation[MAXIV]{ MAX IV Laboratory Lund University Sweden}
\author{Felix Roosen-Runge} 
\affiliation[Lund]{ Division of Physical Chemistry, Lund University, Naturvetarvägen 22, 22100 Lund, Sweden}
\author{Fivos Perakis} 
\affiliation[ Stockholm University]{Department of Physics, AlbaNova University Center, Stockholm University, 10691 Stockholm, Sweden}
\email{f.perakis@fysik.su.se}

\title{Kinetics of ferritin crystal formation and melting in acoustically levitated droplets} 

\begin{document}

\begin{tocentry}
\centering
\includegraphics[height=1.75in,width=3.25in]{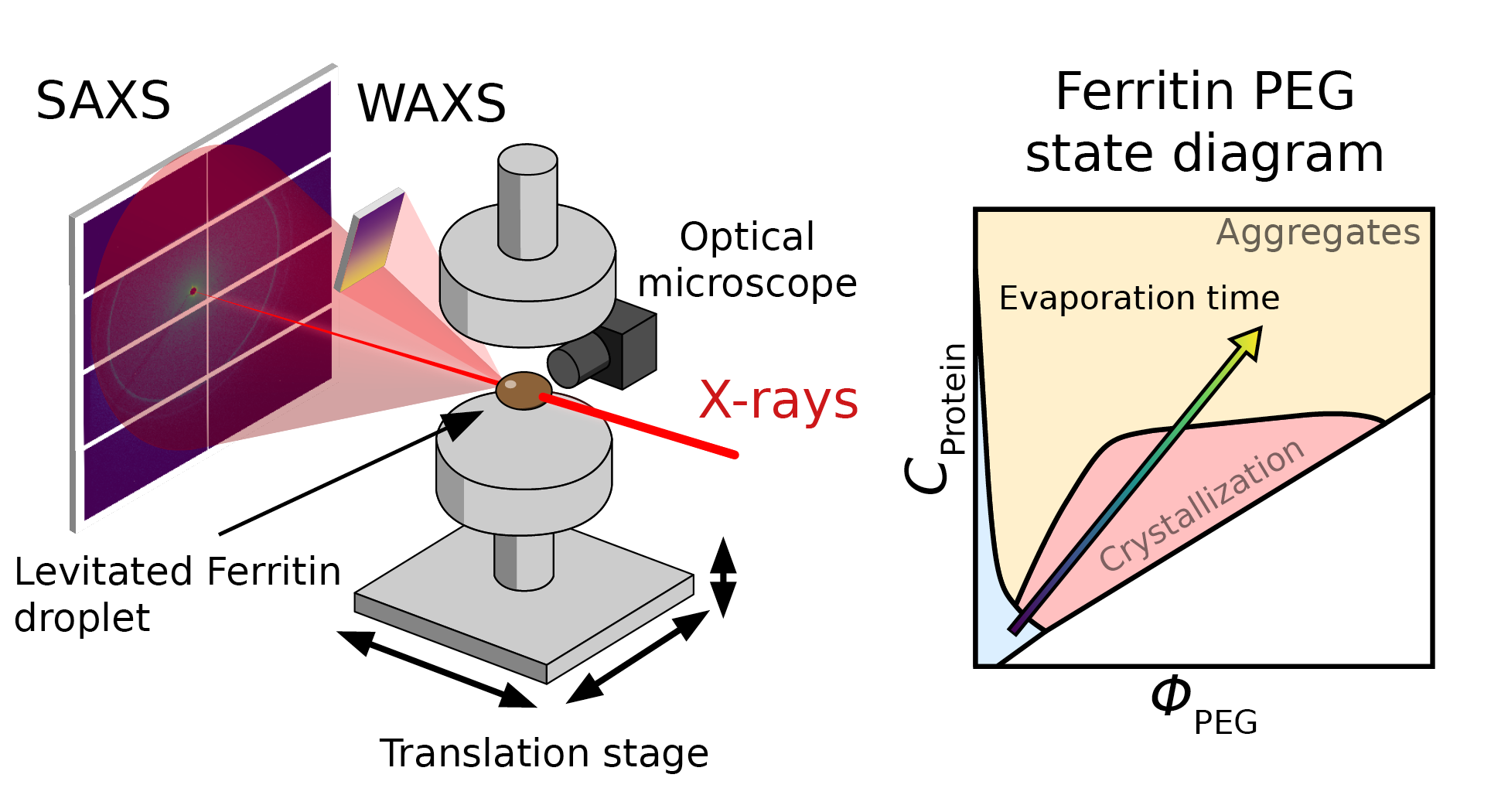}
\end{tocentry}

\begin{abstract}
Understanding protein crystallization pathways is essential for controlling crystallization in structural biology, materials science, and pharmaceutical applications. Classical nucleation theory does not fully capture crystallization processes for several proteins, including ferritin. Here, we combine acoustic levitation with small- and wide-angle X-ray scattering (SAXS and WAXS) to monitor ferritin crystallization in evaporating aqueous polyethylene glycol (PEG) solutions. Acoustic levitation rapidly drives the droplets through a broad range of protein and polymer concentrations, enabling time-resolved measurements of crystallization during evaporation. The scattering data show that ferritin crystals form during evaporation and subsequently lose their crystalline order upon further dehydration. Varying the PEG molecular weight switches between distinct crystallization pathways: one dominated by attractive protein–protein interactions and another dominated by repulsive interactions and excluded-volume effects.
Furthermore, we find that lower-molecular-weight PEG (1000~g/mol) suppresses the dehydration-induced loss of crystalline order observed for higher-molecular-weight PEG (6000~g/mol), providing a simple strategy for improving protein crystal stability.
\end{abstract}

\section*{Introduction}

Protein crystallization underpins structural biology and numerous industrial separation processes~\cite{chen_pharmaceutical_2011}, yet predicting and controlling crystallization pathways remains a major challenge~\cite{holcomb_protein_2017}. 
Traditional models have largely relied on classical nucleation theory\cite{sear_nucleation_2007}. Classical nucleation theory describes crystallization as a stochastic process in which individual monomers cluster until a critical nucleus forms, after which crystal growth proceeds. Although this framework has provided useful information, it has proven insufficient to explain the full range of crystallization phenomena observed in proteins \cite{sear_nucleation_2007,sauter_real-time_2015}. Increasing evidence points to alternative non-classical pathways that involve metastable intermediate phases, liquid–liquid phase separation, or amorphous precursors \cite{gebauer2014pre,sauter_real-time_2015,galkin_are_2000}. Such processes can strongly influence nucleation rates, polymorph selection, and ultimately the quality of the resulting crystals.

Ferritin, an iron-storage protein with a highly symmetric cage-like architecture, has emerged as a particularly intriguing case: Recent studies have demonstrated that ferritin undergoes a non-classical crystallization pathway when crystallized in the presence of CdCl$_2$ \cite{yau2000quasi,houben_mechanism_2020}. Instead of following a two-step nucleation model, ferritin molecules first assemble into amorphous aggregates\cite{houben_mechanism_2020}, which then undergo structural rearrangements until a crystalline order is reached.

Ferritin has been reported to crystallize not only in the presence of CdCl$_2$, but also in the presence of polyethylene glycol (PEG)\cite{tanaka_protein_2002}. This polymer is widely used as a crystallizing agent \cite{mcpherson1976crystallization,rupp2009biomolecular}, as it induces liquid-liquid phase separation \cite{annunziata_observation_2003} via depletion interactions \cite{kulkarni_depletion_2001} which can enhance crystallization \cite{ten_wolde_enhancement_1997}, and is associated with a two-step crystallization mechanism in which the crystals form at the expense of liquid-like aggregates\cite{zhang_role_2012}. 

To probe ferritin crystallization in~situ, acoustic levitation provides a powerful container-less method for studying nucleation processes at high supersaturation \cite{agthe_following_2016,cao_rapid_2012,sonderby_concentrated_2020}. This approach eliminates heterogeneous nucleation from container walls and enables the use of minimal protein volumes while accessing a wide range of concentrations during the evaporation process. 

In this work, we used acoustically levitated droplets to monitor the kinetic pathways of ferritin aggregation, crystal nucleation, and crystalline ordering under controlled conditions, varying PEG concentration and molecular weight (PEG1000, with 1000 g/mol, and PEG6000 with 6000 g/mol). By combining Small-Angle X-ray Scattering (SAXS), Wide-Angle X-ray Scattering (WAXS), and optical microscopy, we captured the evolution from dispersed proteins through concentrated intermediate states to crystalline order, and mapped the interplay between crowding, protein–protein interactions, and dehydration. This approach enables the estimation of a state diagram of ferritin crystallization and provides direct insight into the mechanisms of crystallization.

\section*{Results and Discussion}
Droplets of protein solution with radius $\approx1\unit{\milli\meter}$ (with volume $\approx 10\unit{\micro\liter}$) were injected into an ultrasonic acoustic levitator and simultaneously probed with X-ray scattering and optical microscopy. The volume of the droplets was estimated from time-resolved optical images, while the molecular structure of the protein and water were monitored through X-ray scattering. 

\begin{figure}[h]
    \centering
    \includegraphics[width=1\linewidth]{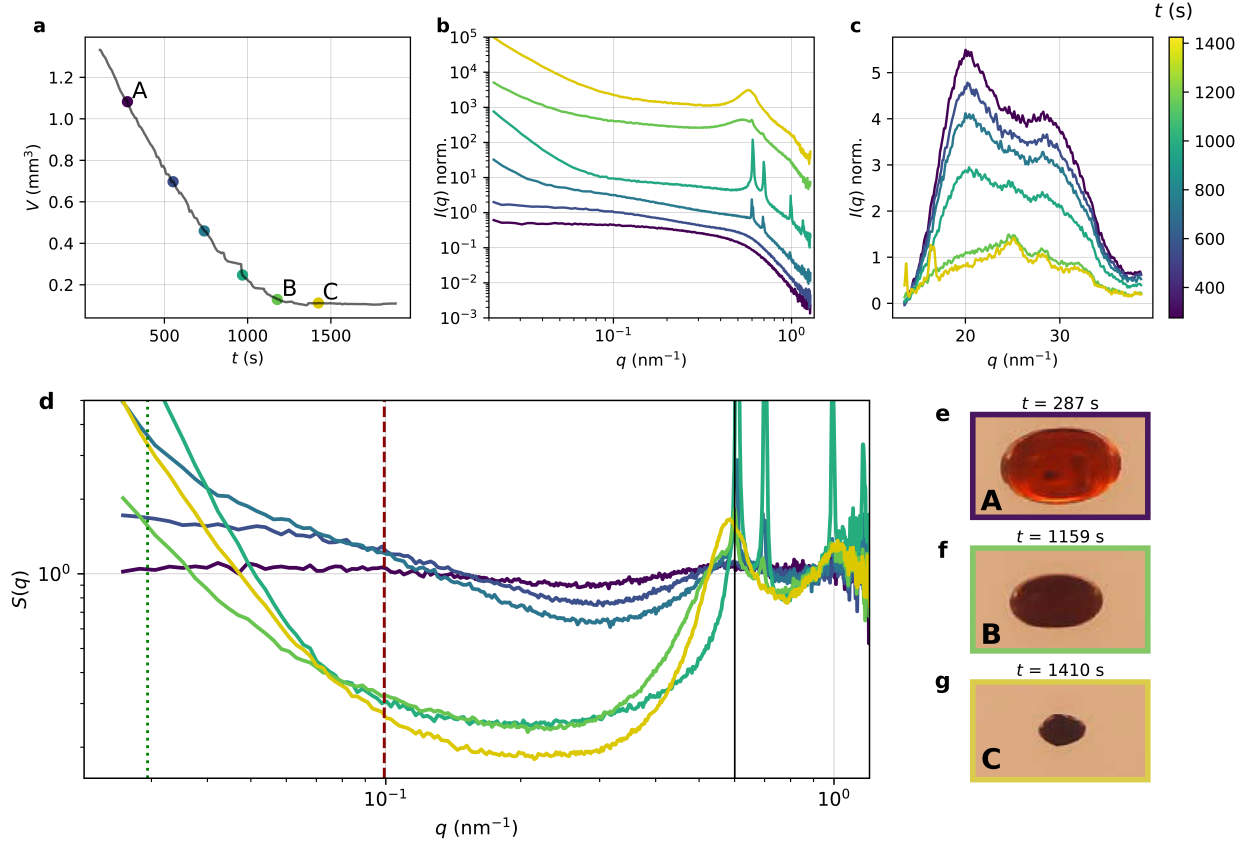}
    \caption{Time evolution of an acoustically levitated ferritin droplet during evaporation. The initial solution contained 50 mg/mL ferritin, 3\% (w/v)  PEG6000, and 150 mM NaCl. (a) Droplet volume as a function of evaporation time $t$. 
    b) SAXS intensity profiles as a function of momentum transfer $q$ at different evaporation times $t$. The curves are shifted for clarity. c) Corresponding WAXS intensity profiles. d) Protein structure factor obtained by dividing the SAXS intensity by the ferritin form factor. e-g) Optical microscopy images corresponding to the labeled time points in panel (a). Colors indicate evaporation time and are shared across panels (a–d). }
    \label{fig:Iq}
\end{figure}

Figure~\ref{fig:Iq}a shows the volume of a droplet of ferritin and polyethylene glycol (PEG) in saline solution (initial concentrations of $c_i^{\mathrm{protein}}=50$~mg/mL Ferritin, $c_i^{\mathrm{PEG}}=3$ w/v \% PEG6000 and 150 mM NaCl) as a function of evaporation time $t$ (Fig.~\ref{fig:Iq}a). Representative microscopy images from which the droplet volume was estimated are shown in Fig.~\ref{fig:Iq}e-g. During the collection of the microscopy images, the X-ray signal was collected by two detectors to cover both SAXS and WAXS ranges. 

Representative SAXS and WAXS curves are shown in Fig.~\ref{fig:Iq}b and Fig.~\ref{fig:Iq}c respectively. The colors from blue to yellow indicate different evaporation times at which they were collected, with $t=0$~s being the time at which the droplet is injected into the levitator. The corresponding estimated volumes are shown in Fig.~\ref{fig:Iq}a as circles with corresponding colors. In Fig.~\ref{fig:Iq}a, the estimated volume decreases linearly with time in the first 600 s, and after a turning point at 1100 s, it becomes almost constant. 

At early times (Fig.~\ref{fig:Iq}b, dark violet curve), the SAXS profile is close to that of ferritin single proteins in solution. At intermediate evaporation times, the SAXS intensity increases at low $q$, while at $t=700$~s protein Bragg peaks appear at $q\approx0.6$~nm$^{-1}$, $q\approx0.7$~nm$^{-1}$, $q\approx1.0$~nm$^{-1}$, as well as additional smaller peaks at higher $q$. Interestingly, from $t\approx1200$~s the height of the Bragg peaks decreases, until $t\approx1400$~s when they are absent. In the WAXS signal (Fig.~\ref{fig:Iq}c), in the earlier times the main contribution is due to two water peaks at $q\approx20$~nm$^{-1}$ and $q\approx28$~nm$^{-1}$. With increasing time $t$, the water peaks decrease in height as the water evaporates, which is also evident by the volume changes with time $t$ in Fig.~\ref{fig:Iq}a. 
At later times, two additional sharp peaks appear at $q\approx13.7$~nm$^{-1}$ and 
$q\approx16.5$~nm$^{-1}$, which correspond to Bragg peaks from PEG semi-crystals \cite{kuttich_x-ray_2020}. At higher $q$-values, the curve presents three peaks at $q\approx25,\,28$ and $32$~nm$^{-1}$. We attribute these peaks to ferrihydrite (6LFh) contained in the protein core together with magnetite and hematite\cite{galvez_comparative_2008,janney_transmission_2000,cowley_structure_2000}.

The experimental protein structure factor $S(q)$, shown for representative conditions in Fig.~\ref{fig:Iq}d and for the different sample compositions in the SI, Fig.~S7, was estimated by dividing the scattering intensity $I(q)$ by the form factor $P(q)$.
$P(q)$ was extracted by measuring a dilute protein solution (with a concentration of 10 mg/mL) in a capillary; see Supplementary Information section "Ferritin Form Factor" for more information. The signal was normalized to ensure that the oscillations of the structure factor at large $q$ were around 1, i.e. $S(q\rightarrow\infty)=1$. 

\begin{figure}[h]
    \centering
    \includegraphics[width=.7\linewidth]{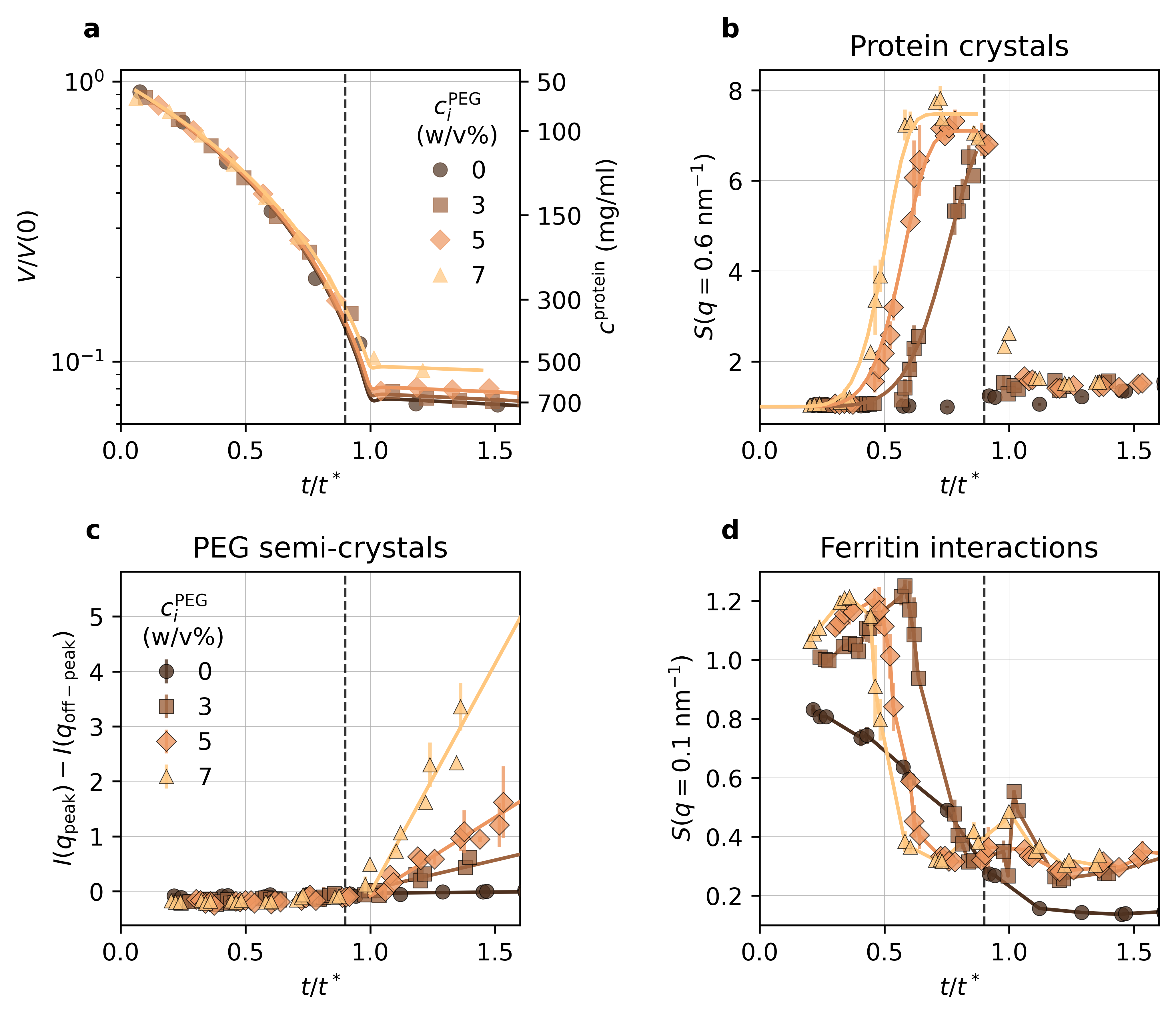}
    \caption{Evaporation kinetics for droplets with different initial PEG concentrations. Initial sample composition is 50~mg/mL ferritin, PEG6000, and 150~mM NaCl, with initial PEG concentrations of 0, 3, 5, and 7 wt\%. (a) Normalized droplet volume (left axis) and corresponding average protein concentration (right axis) as a function of normalized evaporation time $t/t^{*}$. (b) Amplitude of the ferritin Bragg peak, estimated from the structure factor at $q=0.6$~nm$^{-1}$, used to quantify crystalline order.  (c) Amplitude of the PEG semi-crystalline Bragg peak, estimated as $I(q_{\mathrm{peak}})-I(q_{\mathrm{off-peak}})$ with $q_{\mathrm{off-peak}}=14$~nm$^{-1}$ and $q_{\mathrm{peak}}=13.7$~nm$^{-1}$. (d) Structure factor at $q=0.1$~nm$^{-1}$, which is sensitive to protein–protein interactions on length scales of approximately 60~nm.
    }
    \label{fig:summary_concentrations}
\end{figure}

To understand the kinetics of ferritin crystallization, we need to monitor the different components of the solution: water, proteins in solution, proteins in crystals, and PEG. The average ferritin concentration $c^{\mathrm{protein}}$ in the droplet can be estimated from the changes in droplet volume with time. Fig.~\ref{fig:summary_concentrations}a on the left-hand side y-axis shows the time dependence of the normalized volume $V(t)/V(t=0)$ for samples at different initial concentrations of PEG ($c_i^{\mathrm{PEG}}=$~0, 3, 5, 7~w/v~\% ). The right-hand side y-axis shows the corresponding average protein concentration in the droplet. To compare similar evaporation stages (that is, comparable protein concentration), time $t$ was normalized to $t^{*}$, which is the turning point where volume changes are minimal. Before $t^{*}$, the volume changes follow a $\mathcal{D}^2$-law \cite{mcgaughey2002temperature, dallaBarba_revisiting_2021}, which states that the square of diameter $\mathcal{D}$ changes linearly with time. This master curve implies that, at the same normalized time, the droplets have the same protein concentration. After $t^{*}$, the volume decreases linearly with time with a shallow slope, implying that the volume (and therefore the protein concentration) is almost constant. The final protein concentration  depends on $c_i^{\mathrm{PEG}}$, with larger values of $c^{\mathrm{protein}}$ for lower $c_i^{\mathrm{PEG}}$. 

To monitor the presence of crystals, we estimate the amplitude of the protein crystals as $S(q=$~0.6~nm$^{-1}$), i.e. the structure factor at the first protein Bragg peak (Fig.~\ref{fig:Iq}b). The amplitude increases at earlier normalized times for a higher initial PEG concentration $c_i^{\mathrm{PEG}}$ (Fig.~\ref{fig:summary_concentrations}b). In contrast, the amplitude rapidly decreases at $t/t^{*}\geq0.9$ (indicated by the black dashed line) for all $c_i^{\mathrm{PEG}}>0$. This indicates a loss of crystalline order once the protein concentration reaches approximately $c^{\mathrm{protein}}\approx$~400-500~mg/mL.

The behavior of PEG in the solution is analyzed in Fig.~\ref{fig:summary_concentrations}c. The WAXS intensity at $q_{\mathrm{peak}}=13.7$~nm$^{-1}$ provides information on the crystalline ordering of PEG that develops at low hydration. In semicrystalline PEG, polymer chains partially fold into crystalline lamellae separated by amorphous regions~\cite{kuttich_x-ray_2020}. The molecular ordering within the crystalline regions gives rise to a Bragg peak at $q_{\mathrm{peak}}$. To separate the crystalline PEG contribution from the underlying water scattering, its amplitude was estimated as $I(q_{\mathrm{peak}})-I(q_{\mathrm{off-peak}})$ with $q_{\mathrm{off-peak}}=14$~nm$^{-1}$. Fig.~\ref{fig:summary_concentrations}c shows that this amplitude is constant around 0 until time $t/t^{*} \approx 1$. Afterwards, it increases linearly. This indicates that PEG molecules do not form semi-crystals up to the time when the volume is almost constant, which is after the protein crystals have melted. These observations suggest that PEG semi-crystallization is unlikely to be responsible for the loss of crystalline order.

To understand the mechanism of formation of ferritin crystals, protein interactions were estimated as the structure factor at $q=0.1$~nm$^{-1}$. 
Further details of the SAXS analysis and representative fits are provided in the Supplementary Information (Fig. S8). At this $q$, the length-scale probed is around 60~nm, i.e. corresponding to the size of several proteins ($2R_{\mathrm{protein}}\approx12$~nm). Attractive protein–protein interactions result in $S(q=0.1$~nm$^{-1})>1$, while repulsive interactions in $S(q=0.1$~nm$^{-1})<1$. In Fig.~\ref{fig:summary_concentrations}d, $S(q=0.1$~nm$^{-1})$ increases linearly from 1 for all $c_i^{\mathrm{PEG}}>0$. It reaches its maximum with a value of $S(q=0.1$~nm$^{-1})$ at the same time that protein crystallization starts. 
The attractive protein–protein interactions observed before crystallization ($S(q=0.1$~nm$^{-1})>1$) indicate that direct interparticle attractions dominate the crystallization pathway under these conditions.

\begin{figure}[h]
    \centering
    \includegraphics[width=.6\linewidth]{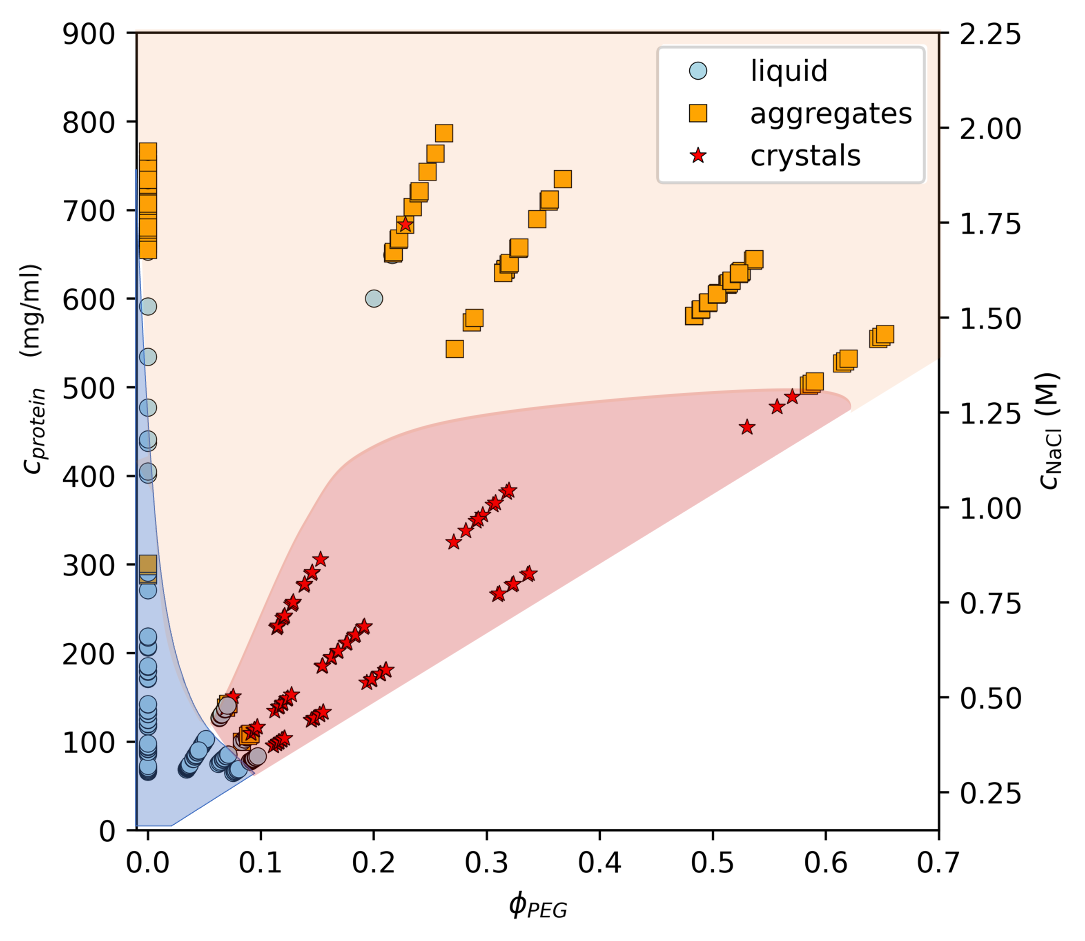}
 \caption{Experimental crystallization state diagram constructed from the evaporation experiments shown in Fig.~\ref{fig:summary_concentrations}. The diagram is a function of PEG volume fraction $\phi_{PEG}$ (for PEG6000) and protein concentration in the droplet $c_{\mathrm{protein}}$. The blue circles represent the solution when mainly monomers are present, while the red stars correspond to solutions with $S(q=0.6$~nm$^{-1})>2$. Orange symbols correspond to samples exhibiting large aggregates (quantified by $S(q=0.03$~nm$^{-1})>1.5$ and $S(q=0.03$~nm$^{-1})>S(q=0.1$~nm$^{-1})$), but no protein Bragg peaks.}
    \label{fig:state-diagram}
\end{figure}

By combining the results collected at different concentrations of PEG6000, we can construct an experimental crystallization state diagram with polymer and protein concentration as parameters (see Fig.~\ref{fig:state-diagram}). When comparing this state diagram with a typical protein crystallization phase diagram \cite{asherie_protein_2004}, it is important to remember that the state diagram we present here does not reflect the equilibrium state of the solution but instead depends on the kinetics of droplet evaporation. Nevertheless, from these results we can infer important information: the solution in the liquid area (light-blue) most likely corresponds to either undersaturated solutions or supersaturated solutions outside the nucleation regime, since no signature of crystals is visible. The crystallization region (red) indicates that, at this position, the sample is either within the nucleation zone or in the precipitation zone, where crystals are present as a result of having formed earlier. The orange region corresponds to droplets that are in the precipitation zone.

\begin{figure}[h]
    \centering
    \includegraphics[width=0.65\linewidth]{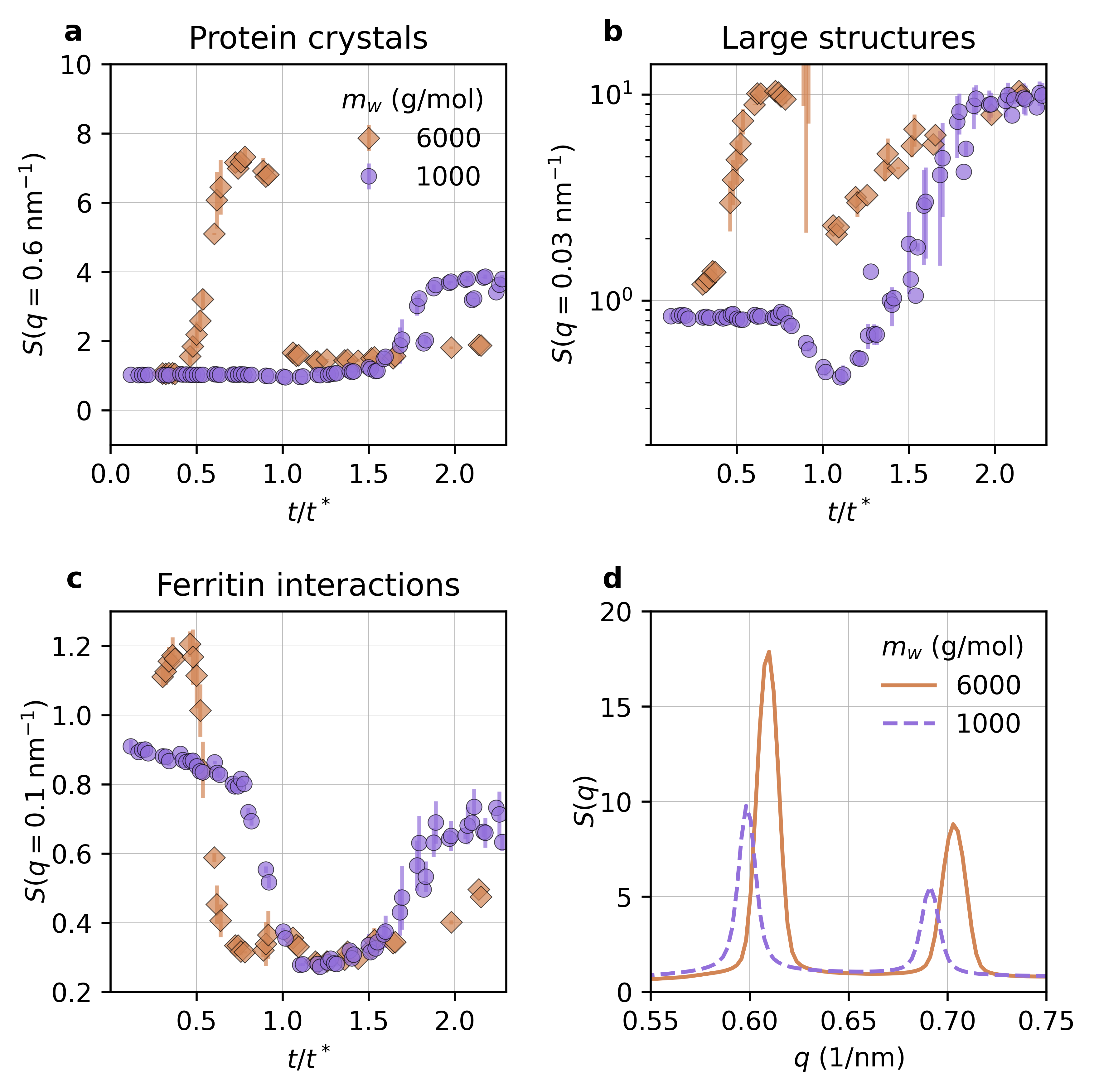}
    \caption{Comparison between droplets initially containing 50 mg/mL ferritin, 5\% (w/v) PEG, and 150 mM NaCl, using either PEG1000 or PEG6000. (a) Ferritin crystal amplitude, estimated from the structure factor at the first Bragg peak $q=0.6$~nm$^{-1}$. (b) Structure factor at $q=0.03$~nm$^{-1}$, sensitive to mesoscale protein organization on length scales of approximately 200 nm. (c) Structure factor at $q=0.1$ nm$^{-1}$ probing protein-protein interactions. (d) Average protein structure factor in the crystalline state for PEG1000 and PEG6000. }
    \label{fig:kinetics_pegsize}
\end{figure}

To better understand the role of PEG in protein crystallization, we repeated the experiment using a different PEG molecular weight.. Fig.~\ref{fig:kinetics_pegsize} compares solutions containing PEG1000 and PEG6000 at the same initial PEG concentration, $c_i^{\mathrm{PEG}}=5$ \% (w/v). The results indicate clear differences, the first of which is the crystallization time: In Fig.~\ref{fig:kinetics_pegsize}a, one can see that for PEG1000, crystallization begins later compared to PEG6000. 
Furthermore, for PEG1000, crystallization occurs only after the droplet has reached the low-water, nearly constant-volume regime, and the crystal amplitude subsequently remains stable within the experimental time window. Thus, in contrast to PEG6000, PEG1000 not only prevents the loss of crystalline order but supports crystal formation under these highly concentrated conditions.

The structure factor at low scattering vector ($q=0.03$~nm$^{-1}$) is sensitive to length scales around 210~nm. Figure~\ref{fig:kinetics_pegsize}b shows that for PEG6000, $S(q=0.03$~nm$^{-1})$ increases around $t/t^*=0.4$, similarly to the growth of the amplitude of the protein crystals. This indicates that the observed increase primarily reflects crystal formation. For PEG1000, the amplitude of protein crystals increases around $t/t^*=1.5$, i.e. after the decrease and subsequent increase of  $S(q=0.03$~nm$^{-1})$ at $t/t^*\approx1.1$. 
For PEG1000, $S(q=0.03\,\mathrm{nm}^{-1})$ remains nearly constant up to $t/t^*\approx0.7$, indicating that no large-scale structures have yet formed. Interestingly, also the protein interactions remain predominantly repulsive throughout the entire process, i.e. $S(q=0.1$~nm$^{-1})<1$. The pronounced change in the structure factor around $t/t^*\approx0.7$ may reflect the multicomponent nature of the effective interactions at high concentrations, where confinement of PEG to the increasingly restricted void space between proteins could enhance the effective repulsion.

Beyond this point, the evolution of the low-$q$ scattering, together with $S(q=0.1$~nm$^{-1})<1$, indicates the formation of larger-scale structures despite the predominantly repulsive protein–protein interactions. Rather than attraction-driven aggregates, these structures may therefore reflect increasingly crowded or jammed protein configurations as the droplet approaches the highly concentrated state in which crystallization occurs. These observations suggest that reducing the PEG molecular weight changes not only the dominant protein interactions but also the kinetic pathway leading to crystallization, with increasingly crowded protein configurations developing before crystal nucleation.

The structure factor of the solutions containing crystals for the two molecular weights are shown in Fig.~\ref{fig:kinetics_pegsize}d. The Bragg peaks for PEG1000 are shifted to lower $q$ values, indicating a larger lattice spacing.
The predominantly repulsive protein--protein interactions observed for PEG1000 suggest that the crystallization pathway differs fundamentally from that observed for PEG6000. For PEG1000, the value of $S(q=0.1$~nm$^{-1})<1$ indicates that attractive protein--protein interactions are not the dominant driving force for crystallization. Instead, the combination of repulsive interactions, increasing crowding, and crystallization only at high protein concentrations is consistent with a pathway dominated by excluded-volume effects, analogous to hard-sphere crystallization~\cite{pusey1986phase,frenkel1999entropy}. In this picture, the evolution of the low-$q$ scattering prior to crystallization may reflect increasingly crowded or jammed protein configurations rather than attraction-driven aggregates.

An important parameter to consider when comparing protein solutions with different PEG molecular weights is the ratio $\xi = R_g/R_{\mathrm{protein}}$ between the radius of gyration, $R_g$, of PEG and the radius of protein $R_{\mathrm{protein}}$ \cite{tanaka_protein_2002,ilett_phase_1995}. This parameter can strongly influence the crystallization phase diagram by shifting the solubility line, precipitation and nucleation regions, giving also rise to a liquid-liquid phase separation region.  Liquid-liquid phase separation  is present in the phase diagram for $\xi>0.25$  \cite{tanaka_protein_2002,ilett_phase_1995}. The two different molecular weights shown correspond to $\xi(\mathrm{PEG1000})=0.16$ and $\xi(\mathrm{PEG6000})=0.4$. The two values of $\xi$ imply that only for PEG6000 the solution could be close to a liquid-liquid phase separation boundary, which is known to enhance protein crystallization via non-classical pathways \cite{ten_wolde_enhancement_1997}. This interpretation is also consistent with previous experiments on ferritin–PEG mixtures, which showed that, for a given protein concentration, the PEG concentration required to induce crystallization decreases with increasing PEG molecular weight~\cite{tanaka_protein_2002}. However, it is not clear why     no loss of crystalline order is observed for PEG1000.

A likely explanation for the loss of crystalline order is dehydration. Protein crystals are made up of 30~-~70\% (in volume) of  water \cite{carugo_protein_2017}, which is necessary for the presence of hydrogen bonds that stabilize interactions between proteins \cite{schoenborn_hydration_1995}. Hence, if the water content decreases below a certain threshold, the crystal lattice may become unstable.

The hypothesis that the loss of crystalline order is caused by dehydration is supported by the evolution of the Bragg peak positions. Before the crystals lose their crystalline order, their Bragg peak positions shift to higher $q$ values (see SI, Fig.~S3), corresponding to a smaller unit cell, as expected for crystal dehydration\cite{einstein_insulin_1962}. The Bragg peaks disappear only shortly before $t^*$, when the volume is almost constant, possibly due to a liquid-solid transition, which could destabilize the protein crystal.  
However, this transition does not stop the formation of protein crystals when the solution contains PEG1000.  The amount of water, estimated from $I(q=20 $~nm$^{-1})$, reaches similar values for  PEG6000 and PEG1000 (see SI, Fig.~S5), but protein crystals lose their crystalline order within the experimental time for PEG6000, whereas they form and remain crystalline under similarly low-water conditions for PEG1000.

A possible explanation for the difference in protein crystal stability between PEG1000 and PEG6000 is that PEG1000 adsorbs onto proteins within the crystal, thereby retaining a higher local water content. PEG has previously been shown to adsorb onto proteins through hydrophobic interactions and hydrogen bonding~\cite{wu_binding_2014,bekale_role_2015} and can also be incorporated into protein crystals~\cite{mcpherson2026note}.
The presence of PEG within the crystal lattice could explain both the larger lattice spacing and the enhanced stability of PEG1000 crystals despite similar overall hydration levels measured by WAXS.

One possible interpretation is that, for PEG6000, protein crystallization begins before significant PEG adsorption occurs. As dehydration proceeds, the crystals gradually lose stability and eventually lose their crystalline order. In contrast, for PEG1000, the droplet reaches a lower hydration level before crystallization begins, allowing PEG to adsorb onto the protein surface. PEG molecules associated with the proteins may then stabilize the crystal by helping to retain water within the crystal lattice.

\section*{Conclusions}

We have demonstrated that acoustic levitation combined with time-resolved SAXS/WAXS provides a powerful approach for following protein crystallization in~situ using only microliter sample volumes. By continuously concentrating the sample during evaporation, this method enables rapid mapping of crystallization pathways and state diagrams while requiring only minimal amounts of protein. 

Our results reveal that the molecular weight of PEG not only controls the onset of crystallization but also alters the underlying crystallization mechanism. For PEG6000, attractive protein–protein interactions develop before crystallization, indicating a pathway dominated by direct interparticle attractions. In contrast, PEG1000 gives rise to predominantly repulsive protein–protein interactions and crystallization only at high protein concentrations, consistent with a distinct pathway governed by crowding and excluded-volume effects. While crystals formed with PEG6000 lose their crystalline order at late evaporation stages, crystals with PEG1000 form under highly concentrated, low-water conditions and remain stable throughout the experimental time window despite comparable overall hydration levels.
This observation suggests that shorter PEG chains modify the local environment of the protein crystals, providing enhanced resistance against dehydration-induced loss of crystalline order.

More broadly, these results demonstrate that polymer molecular weight can be used to tune not only crystallization kinetics but also the crystallization pathway and the stability of the resulting crystals. More generally, acoustic levitation combined with time-resolved X-ray scattering enables rapid, low-volume studies of protein crystallization under continuously evolving solution conditions, providing new opportunities to investigate non-classical crystallization mechanisms and optimize crystallization protocols.

\section*{Experimental}
\subsection*{Small- and Wide-Angle X-ray Scattering Experimental parameters}
The experiment was performed at the CoSAXS beamline (MAX~IV in Lund, Sweden). The size of the X-ray beam  was $100 \times 100$~\unit{\micro\meter} and the photon energy was set to 18 keV. Measurements were acquired at multiple positions across the droplet, after which only those corresponding to the central region were retained for quantitative analysis. The exposure time was 0.1~s, with an average interval of 1.4~s between successive measurements, determined by the motor movement between positions. The X-rays were collected by the two detectors EIGER2~4M (SAXS) and Mythen~100K (1D WAXS) at a distance of 8.065~m and 0.195~m respectively. These parameters correspond to a $q$-range of 0.03-1~nm$^{-1}$ and 15-35~nm$^{-1}$.

The intensity $I(q)$ in the SAXS range was obtained via:

\begin{equation}
    I(q)=\left(I_{\mathrm{raw}}(q)-I_{\mathrm{bkg}}(q)\right)\left(-\log\left(\frac{I_0^{\mathrm{sample}}T^{\mathrm{ empty}}}{T^{\mathrm{sample}}I_0^{\mathrm{ empty}}}\right)\right),
    \label{eq:I_exp}
\end{equation}

 where $I_{\mathrm{raw}}(q)$ is the azimuthally integrated detector intensity,  $I_{\mathrm{bkg}}(q)$ is the background measured from a buffer droplet, $T$ is the measured transmission and $I_0$ the corresponding incident beam intensity used for transmission normalization. \cite{sonderby_concentrated_2020}. 

The intensity in the WAXS range was obtained from Mythen, which is a 1D detector, and hence no azimuthal integration was necessary. The experimental scattering intensity was extracted using equation \ref{eq:I_exp},  with $I_{\mathrm{bkg}}(q)$ being the scattering measured without a droplet present in the beam path. 

The intensities used for the analysis were the intensities from measurements in the center of the droplet. A description of the estimation of the center can be found in the Supplementary Information section Droplet center definition. Only measurements acquired near the droplet center were used for quantitative analysis in order to minimize concentration gradients arising during drying and to probe the bulk of the droplet (see Supplementary Information section Homogeneity of droplets for more information).

\subsection*{Ultrasonic acoustic levitator}
The solution was levitated using an Ultrasonic Acoustic levitator (Tec5 AG, Germany) operating at a frequency of 100~kHz. The distance between the transducer and the reflector was around 7.5~mm. Protein droplets of about 1~mm in radius were injected with a Hamilton syringe.

\subsection*{Volume Estimations}
The images were recorded with a microscope positioned perpendicularly to the X-ray beam. Each droplet image was fitted with an ellipse to determine the droplet height $h$ and width $w$ of the droplet. Assuming rotational symmetry about the vertical axis, the volume was calculated as $V=\frac{\pi}{6} hw^2$.

During the initial evaporation stage ($t<t^*$) the droplet radius followed the classical $\mathcal{D}^2$-law for which the volume changes according to the following equation:
\begin{equation}
    r^2 = r_0^2-Kt,
\end{equation}
with $r_0$ being the radius at time $t=0$~s (i.e. when the droplet was injected) and $K$ a constant as shown in SI, Fig.~S2.

For time $t>t^*$ the volume was approximated by a linear function of time. The extracted value of $t^*$ was subsequently used to normalize the time axis in Figs.\ref{fig:summary_concentrations} and \ref{fig:kinetics_pegsize}.

\subsection*{Sample preparation}
Commercial horse-spleen ferritin (Sigma-Aldrich F4503) was supplied as a stock solution containing  71~mg/mL protein in 150~mM NaCl. The PEG powder (PEG1000 Sigma-Aldrich  8.07488, PEG6000 Sigma-Aldrich  8.07491) was dissolved in aqueous solution with 150~mM NaCl. 
The PEG solution was then mixed with the ferritin stock solution to obtain the desired final protein and PEG concentrations. The compositions of the measured samples are listed in Table \ref{tab:samples}.

\begin{table}[tbhp]
 \centering
 \caption{\textbf{Sample compositions used in this study, including protein concentration $c$, PEG concentration $c_{\mathrm{PEG}}$, PEG molecular weight $m_w$ and size ratio  $\xi=R_g/R_{\mathrm{protein}}$. The NaCl concentration was 150~mM for all solutions. }}
 \label{tab:samples}
 \begin{tabular}{lccc}
 \hline
 $c$ (mg/mL) &$c_{\mathrm{PEG}}$ (w/v\%) & $m_w$ (g/mol)  & $\xi=R_g/R_{\mathrm{protein}}$ \\
 \hline
 50 &2  & 6000  & 0.40\\
 50 &3  & 6000 & 0.40\\
 50 &5  & 6000 & 0.40\\
 50 &7  & 6000 & 0.40\\
 50 &5  & 1000 & 0.16 \\
 \hline
 \end{tabular}
\end{table}

\section*{Acknowledgements}

We acknowledge MAX~IV in Lund for provision of beamtime at the CoSAXS instrument and would like to thank the staff for their assistance. The data presented here are collected as part of measurement time awarded to proposal numbers 20241728 and 20250036. In addition, we acknowledge HPC clusters at MAX IV for providing computer resources to perform the analysis. 
FP acknowledges financial support by the Swedish National Research Council (Vetenskapsrådet) under Grant No. 2019-05542, 2023-05339 and within the Röntgen-Ångström Cluster Grant No. 2019-06075, and the financial support from Knut och Alice Wallenberg foundation (WAF, Grant. No. 2023.0052). This research is supported by the Wenner-Gren Foundations (Project No. UPD2021-0144). LB
acknowledges financial support by the Swedish National Research Council (Vetenskapsrådet)
under Grant No. 2023-05572. F.P., I.A., and A.G. acknowledge funding from the European Union’s Horizon Europe research and innovation program under the Marie Skłodowska-Curie grant agreement No. 101081419 (PRISMAS) (F.P. and I.A.) and 101149230 (CRYSTAL-X) (F.P. and A.G.).

\section*{Author contributions}

AG, IA and FP designed the experiment, along with discussions with LB, TP and FRR. AG, IA prepared and handled the samples. AG, IA, AK, SB, TP, FRR and FP operated CoSAXS and collected data together with the rest of the experimental team. AG and IA performed online data processing and analysis at CoSAXS. AG performed offline data processing and analysis. AG, IA, AK, SB, TP, FRR and FP  performed the experiments and discussed the final results. The manuscript was written by AG with the input of all authors.

\section*{Competing interests}
The authors declare no conflict of interest.

\subsection*{Data availability}
Data are available upon request. 

\subsection*{Code availability}
The code used to analyze the data in this study is available from the corresponding authors upon request.

\bibliography{references}

@article{pusey1986phase,
  title={Phase behaviour of concentrated suspensions of nearly hard colloidal spheres},
  author={Pusey, Peter N and Van Megen, W},
  journal={Nature},
  volume={320},
  number={6060},
  pages={340-342},
  year={1986}
}

@article{frenkel1999entropy,
  title={Entropy-driven phase transitions},
  author={Frenkel, Daan},
  journal={ Phys. A: Stat. Mech. Appl.},
  volume={263},
  number={1-4},
  pages={26--38},
  year={1999},
}

@Article{kuttich_x-ray_2020,
author ="Kuttich, Björn and Matt, Alexander and Appel, Christian and Stühn, Bernd",
title  ="X-ray scattering study on the crystalline and semi-crystalline structure of water/PEG mixtures in their eutectic phase diagram",
journal  ="Soft Matter",
year  ="2020",
volume  ="16",
issue  ="45",
pages  ="10260-10267",
doi  ="10.1039/D0SM01601B",
}

@article{schoenborn_hydration_1995,
title = {Hydration in protein crystallography},
journal = {Prog. Biophys. Mol. Biol.},
volume = {64},
number = {2},
pages = {105-119},
year = {1995},
issn = {0079-6107},
doi = {https://doi.org/10.1016/0079-6107(95)00012-7},
author = {Benno P. Schoenborn and Angel Garcia and Robert Knott}
}

@article{cao_rapid_2012,
    author = {Cao, Hui-Ling and Yin, Da-Chuan and Guo, Yun-Zhu and Ma, Xiao-Liang and He, Jin and Guo, Wei-Hong and Xie, Xu-Zhuo and Zhou, Bo-Ru},
    title = {Rapid crystallization from acoustically levitated droplets},
    journal = {J. Acoust. Soc. Am.},
    volume = {131},
    number = {4},
    pages = {3164-3172},
    year = {2012},
    month = {04},
    issn = {0001-4966},
    doi = {10.1121/1.3688494},
}

@article{agthe_following_2016,
  title={Following in real time the two-step assembly of nanoparticles into mesocrystals in levitating drops},
  author={Agthe, Michael and Plivelic, Tom{\'a}s S and Labrador, Ana and Bergstrom, Lennart and Salazar-Alvarez, German},
  journal={Nano Lett.},
  volume={16},
  number={11},
  pages={6838--6843},
  year={2016},
}

@article{holcomb_protein_2017,
  title={Protein crystallization: Eluding the bottleneck of X-ray crystallography},
  author={Holcomb, Joshua and Spellmon, Nicholas and Zhang, Yingxue and Doughan, Maysaa and Li, Chunying and Yang, Zhe},
  journal={AIMS Biophys.},
  volume={4},
  number={4},
  pages={557},
  year={2017}
}

@article{chen_pharmaceutical_2011,
author = {Chen, Jie and Sarma, Bipul and Evans, James M. B. and Myerson, Allan S.},
title = {Pharmaceutical Crystallization},
journal = {Cryst. Growth Des.},
volume = {11},
number = {4},
pages = {887-895},
year = {2011},
doi = {10.1021/cg101556s},
}

@article{sear_nucleation_2007,
doi = {10.1088/0953-8984/19/3/033101},
url = {https://doi.org/10.1088/0953-8984/19/3/033101},
year = {2007},
month = {1},
volume = {19},
number = {3},
pages = {033101},
author = {Sear, Richard P},
title = {Nucleation: theory and applications to protein solutions and colloidal suspensions},
journal = {J. Phys.: Condens. Matter}
}

@article{cowley_structure_2000,
title = {The Structure of Ferritin Cores Determined by Electron Nanodiffraction},
journal = {J. Struct. Biol.},
volume = {131},
number = {3},
pages = {210-216},
year = {2000},
doi = {https://doi.org/10.1006/jsbi.2000.4292},
author = {J.M. Cowley and Dawn E. Janney and R.C. Gerkin and Peter R. Buseck},
}

@article{mcpherson2026note,
  title={A note on the appearance of PEG in macromolecular crystals},
  author={McPherson, Alexander},
  journal={Acta Crystallogr. F Struct. Biol. Commun.},
  volume={82},
  number={1},
  year={2026},
}

@article{mcgaughey2002temperature,
  title={Temperature discontinuity at the surface of an evaporating droplet},
  author={McGaughey, AJH and Ward, CA},
  journal={J. Appl. Phys.},
  volume={91},
  number={10},
  pages={6406--6415},
  year={2002},
}

@article{janney_transmission_2000,
  title={Transmission electron microscopy of synthetic 2-and 6-line ferrihydrite},
  author={Janney, Dawn E and Cowley, John M and Buseck, Peter R},
  journal={Clays Clay Miner.},
  volume={48},
  number={1},
  pages={111--119},
  year={2000},
}

@article{galvez_comparative_2008,
author = {Gálvez, Natividad and Fernández, Bel{\'e}n and Sánchez, Purificación and Cuesta, Rafael and Ceolín, Marcelo and Clemente-León, Miguel and Trasobares, Susana and López-Haro, Miguel and Calvino, Jose J. and St{\'e}phan, Odile and Domínguez-Vera, Jos{\'e} M.},
title = {Comparative Structural and Chemical Studies of Ferritin Cores with Gradual Removal of their Iron Contents},
journal = {J. Am. Chem. Soc.},
volume = {130},
number = {25},
pages = {8062-8068},
year = {2008},
doi = {10.1021/ja800492z},
}

@article{bekale_role_2015,
title = {The role of polymer size and hydrophobic end-group in PEG–protein interaction},
journal = { Colloids Surf. B},
volume = {130},
pages = {141-148},
year = {2015},
issn = {0927-7765},
doi = {https://doi.org/10.1016/j.colsurfb.2015.03.045},
author = {L. Bekale and D. Agudelo and H.A. Tajmir-Riahi},
}

@Article{wu_binding_2014,
author ="Wu, Jiang and Zhao, Chao and Lin, Weifeng and Hu, Rundong and Wang, Qiuming and Chen, Hong and Li, Lingyan and Chen, Shengfu and Zheng, Jie",
title  ="Binding characteristics between polyethylene glycol (PEG) and proteins in aqueous solution",
journal  ="J. Mater. Chem. B",
year  ="2014",
volume  ="2",
issue  ="20",
pages  ="2983-2992",
doi  ="10.1039/C4TB00253A",
}

@article{sonderby_concentrated_2020,
author = "S{\o}nderby, Pernille and S{\"{o}}derberg, Christopher and Frank{\ae}r, Christian G. and Peters, G{\"{u}}nther and Bukrinski, Jens T. and Labrador, Ana and Plivelic, Tom{\'{a}}s S. and Harris, Pernille",
title = "{Concentrated protein solutions investigated using acoustic levitation and small-angle X-ray scattering}",
journal = "J. Synchrotron Radiat.",
year = "2020",
volume = "27",
number = "2",
pages = "396--404",
month = "3",
doi = {10.1107/S1600577519016977},
}

@article{carugo_protein_2017,
title = {Protein hydration: Investigation of globular protein crystal structures},
journal = {Int. J. Biol. Macromol.},
volume = {99},
pages = {160-165},
year = {2017},
doi = {https://doi.org/10.1016/j.ijbiomac.2017.02.073},
author = {Oliviero Carugo},
}

@article{dallaBarba_revisiting_2021,
    author = {Dalla Barba, F. and Wang, J. and Picano, F.},
    title = {Revisiting D2-law for the evaporation of dilute droplets},
    journal = {Phys. Fluids},
    volume = {33},
    number = {5},
    pages = {051701},
    year = {2021},
    month = {05},
    doi = {10.1063/5.0051078},
}

@article{houben_mechanism_2020,
    title = {A mechanism of ferritin crystallization revealed by cryo-{STEM} tomography},
    volume = {579},
    issn = {1476-4687},
    url = {https://doi.org/10.1038/s41586-020-2104-4},
    doi = {10.1038/s41586-020-2104-4},
    number = {7800},
    journal = {Nature},
    author = {Houben, Lothar and Weissman, Haim and Wolf, Sharon G. and Rybtchinski, Boris},
    month = mar,
    year = {2020},
    pages = {540--543},
}

@article{yau2000quasi,
  title={Quasi-planar nucleus structure in apoferritin crystallization},
  author={Yau, S-T and Vekilov, Peter G},
  journal={Nature},
  volume={406},
  number={6795},
  pages={494--497},
  year={2000},
}

@article{gebauer2014pre,
  title={Pre-nucleation clusters as solute precursors in crystallisation},
  author={Gebauer, Denis and Kellermeier, Matthias and Gale, Julian D and Bergstr{\"o}m, Lennart and C{\"o}lfen, Helmut},
  journal={Chem. Soc. Rev.},
  volume={43},
  number={7},
  pages={2348--2371},
  year={2014},
}

@article{einstein_insulin_1962,
    title = {Insulin. {Some} shrinkage stages of sulfate and citrate crystals},
    volume = {15},
    doi = {10.1107/S0365110X62000079},
    number = {1},
    urldate = {2025-12-04},
    journal = {Acta Crystallogr.},
    author = {Einstein, J. R. and Low, B. W.},
    month = {jan},
    year = {1962},
    pages = {32-34},
}

@book{rupp2009biomolecular,
  title={Biomolecular crystallography: principles, practice, and application to structural biology},
  author={Rupp, Bernhard},
  year={2009},
  publisher={Garland Science}
}

@Article{zhang_role_2012,
author ="Zhang, Fajun and Roosen-Runge, Felix and Sauter, Andrea and Roth, Roland and Skoda, Maximilian W. A. and Jacobs, Robert M. J. and Sztucki, Michael and Schreiber, Frank",
title  ="The role of cluster formation and metastable liquid—liquid phase separation in protein crystallization",
journal  ="Faraday Discuss.",
year  ="2012",
volume  ="159",
issue  ="0",
pages  ="313-325",
publisher  ="The Royal Society of Chemistry",
doi  ="10.1039/C2FD20021J",
url  ="http://dx.doi.org/10.1039/C2FD20021J",
}

@article{mcpherson1976crystallization,
  title={Crystallization of proteins from polyethylene glycol.},
  author={McPherson Jr, ALEXANDER},
  journal={J. Biol. Chem.},
  volume={251},
  number={20},
  pages={6300--6303},
  year={1976},
  publisher={Elsevier}
}

@article{tanaka_protein_2002,
    title = {Protein crystallization induced by polyethylene glycol: {A} model study using apoferritin},
    volume = {117},
    journal = {J. Chem. Phys.},
    author = {Tanaka, Shinpei and Ataka, Mitsuo},
    year = {2002},
    pages = {3504--3510},
}

@article{ilett_phase_1995,
    title = {Phase behavior of a model colloid-polymer mixture},
    volume = {51},
    doi = {10.1103/PhysRevE.51.1344},
    number = {2},
    journal = {Phys. Rev. E.},
    author = {Ilett, S. M. and Orrock, A. and Poon, W. C. K. and Pusey, P. N.},
    year = {1995},
    pages = {1344--1352},
}

@article{ten_wolde_enhancement_1997,
    title = {Enhancement of {Protein} {Crystal} {Nucleation} by {Critical} {Density} {Fluctuations}},
    volume = {277},
    doi = {10.1126/science.277.5334.1975},
    number = {5334},
    journal = {Science},
    author = {ten Wolde, Pieter Rein and Frenkel, Daan},
    year = {1997},
    pages = {1975--1978},
}

@article{sauter_real-time_2015,
    title = {Real-{Time} {Observation} of {Nonclassical} {Protein} {Crystallization} {Kinetics}},
    volume = {137},
    url = {http://dx.doi.org/10.1021/ja510533x},
    doi = {10.1021/ja510533x},
    number = {4},
    journal = {J. Am. Chem. Soc.},
    author = {Sauter, Andrea and Roosen-Runge, Felix and Zhang, Fajun and Lotze, Gudrun and Jacobs, Robert M. J. and Schreiber, Frank},
    month = feb,
    year = {2015},
    pages = {1485--1491},
}

@article{galkin_are_2000,
    title = {Are {Nucleation} {Kinetics} of {Protein} {Crystals} {Similar} to {Those} of {Liquid} {Droplets}?},
    volume = {122},
    url = {http://pubs.acs.org/doi/abs/10.1021/ja9930869},
    number = {1},
    journal = {J. Am. Chem. Soc.},
    author = {Galkin, Oleg and Vekilov, Peter G},
    year = {2000},
    pages = {156--163},
}

@article{asherie_protein_2004,
    title = {Protein crystallization and phase diagrams},
    volume = {34},
    journal = {Methods},
    author = {Asherie, N.},
    year = {2004},
    pages = {266--272},
}

@article{annunziata_observation_2003,
    title = {Observation of liquid-liquid phase separation for eye lens gamma {S}- crystallin},
    volume = {100},
    journal = {Proc. Natl. Acad. Sci. USA},
    author = {Annunziata, O. and Ogun, O. and Benedek, G. B.},
    year = {2003},
    pages = {970--974},
}

@article{kulkarni_depletion_2001,
    title = {Depletion interactions and protein crystallization},
    volume = {232},
    journal = {J. Cryst. Growth},
    author = {Kulkarni, A. M. and Zukoski, C. F.},
    year = {2001},
    pages = {156--164},
}

\end{document}